 \documentclass[%
 reprint,
 amsmath,amssymb,
 aps,
]{revtex4-2}

\usepackage{flushend}
\usepackage{verbatim}
\usepackage{verbatim,upgreek}
\usepackage{chngcntr}
\usepackage{graphicx}
\usepackage{dcolumn}
\usepackage{bm}
\usepackage{comment}
\usepackage{amsmath}
\usepackage{physics}
\usepackage{hyperref}
\hypersetup{
    colorlinks=true,
    citecolor=blue,
    }

\usepackage{color}

\begin{document}

\preprint{APS/123-QED}

\title{A Bayesian Approach towards Atomically-precise Localization in Fluorescence Microscopy}

\author{Yuqin Duan$^{1}$}
\email{sophiayd@mit.edu}
\author{Qiushi Gu$^{1}$}

\author{Hanfeng Wang$^{1}$}
\author{Yong Hu$^{1}$}
\author{Kevin C. Chen$^{1}$}

\author{Matthew E. Trusheim$^{1,2}$}

\author{Dirk R. Englund$^{1}$}
\email{englund@mit.edu}

\affiliation{\\
$^{1}$ Massachusetts Institute of Technology, 50 Vassar Street, Cambridge, MA 02139, USA\\
$^{2}$ DEVCOM, U.S. Army Research Laboratory, Adelphi, MD, USA
}


\begin{abstract}

Super-resolution microscopy has revolutionized the imaging of complex physical and biological systems by surpassing the Abbe diffraction limit. Recent advancements, particularly in single-molecule localization microscopy (SMLM), have pushed localization below nanometer precision, by applying prior knowledge of correlated fluorescence emission from single emitters. However, achieving a refinement from $1$~nm to $1$~\AA ngstr{\"o}m demands a hundred-fold increase in collected photon signal. This quadratic resource scaling imposes a fundamental barrier in SMLM, where the intense photon collection is challenged by photo-bleaching, prolonged integration times, and inherent practical constraints. Here, we break this limit by harnessing the periodic nature of the atomic lattice structure. Applying this discrete grid imaging technique (DIGIT) in a quantum emitter system, we observe an exponential collapse of localization uncertainty once surpassing the host crystal's atomic lattice constant. We further applied DIGIT to a large-scale quantum emitter array, enabling parallel positioning of each emitter through wide-field imaging. Collectively, these advancements establish DIGIT as a competitive tool for achieving unprecedented, precise measurements, ultimately paving the way to direct optical resolution of crystal and atomic features within quantum and biological systems. 


\end{abstract}

\maketitle



\begin{figure*}
\includegraphics[width=1\textwidth]{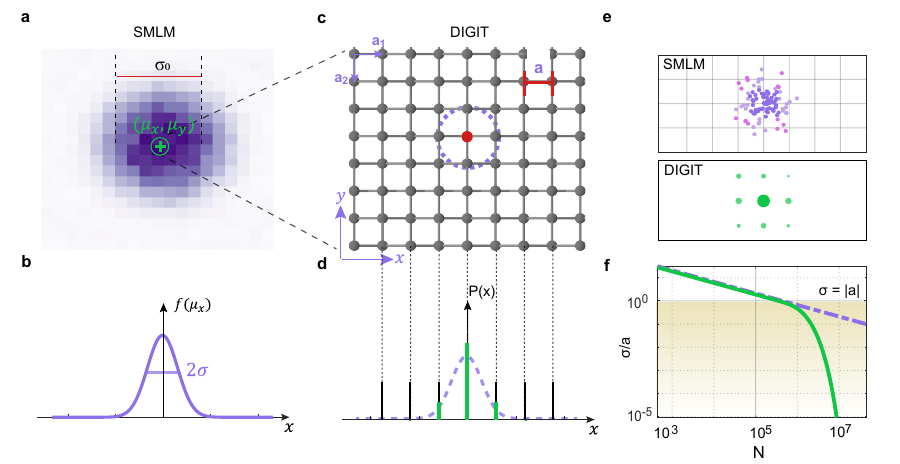}
\caption{\label{fig1}\textbf{DIGIT concept.} \textbf{a,} Farfield PSF of an emitter with a diffraction-limited width $\sigma_0$ and its reconstructed location $(\mu_x,\mu_y)$ using SMLM. \textbf{b,} Probability density function $f(\mu_x)$ for a continuous variable $x$.
\textbf{c,} A zoomed-in view reveals the emitters’ atomic structure, defined by a lattice with constant ($a_1$,$a_2$). Unlike a continuous spatial distribution, the emitters’ positions are restricted to discrete sites determined by the underlying crystal lattice. \textbf{d,} Comparison of posterior localization probability $P(x)$  between conventional SMLM (purple) and DIGIT-equipped SMLM (green). \textbf{e,} An exemplary DIGIT localization in two dimensions. The color coding in SMLM represents distinct lattice groupings. \textbf{f,} Normalized localization precision $\sigma$ as a function of total detected number of photons $N$: conventional SMLM follows the shot noise scaling (purple) while DIGIT exhibits exponential decay (green) when localization approaches lattice constant ($\sigma\sim a$).}
\end{figure*}

\section{Introduction}

Optical super-resolution microscopy has become an important tool for fundamental studies in biology and physics due to its capability to resolve densely packed structures far below the diffraction limit. Recent advancements have enabled lateral resolutions into the sub-nanometer regime localizing fluorescent dyes~\cite{reinhardt2023aangstrom,radmacher2024doubling,pertsinidis2010subnanometre,sahl2024direct}, nanoparticles~\cite{lee2023indefinite} and quantum emitters~\cite{pfender_single-spin_2014, schroder_scalable_2017,rittweger2009sted}. However, pushing localization beyond sub-nanometer precision to atomic scales has remained challenging due to the fundamental shot noise scaling limit that scales inversely with the square root of the photon numbers $N$~\cite{SMIM_2021,scheiderer2025minflux,HellReview2007}. This scaling implies that achieving finer spatial precision requires quadratically increased photon counts, quickly reaching practical limitations such as photon budgets, photobleaching, sample drift and slow noise processes~\cite{Sahl2024}. These constraints significantly restrict real-time tracking of dynamic molecular processes.

The scaling limit arises due to the standard statistics of estimating the mean quantity (e.g. emitter position) over multiple independent measurements. This standard statistical procedure, however, does not capture the physical constraints that are present in many physical systems of interest. For instance, quantum defects reside at discrete lattice sites~\cite{awschalom2018quantum, aharonovich2016solid}, nuclear pores are organized by periodic protein rings~\cite{ibarra2015nuclear,Hampoelz2019}, and synthetic nanostructures follow designed grids~\cite{yim2017significance,PEREZPAGE201651}. Here we introduce a discrete-grid imaging technique (DIGIT) that includes these physical constraints and transforms localization space from continuous to a discrete set of configurations. The localization process then becomes a decision between these emitter configurations, rather than independent estimation of each mean position. This enables the determination of emitter positions with resolution exponentially surpassing the conventional scaling.


We experimentally implement DIGIT to achieve atomic-scale super-resolution imaging of quantum emitters in diamond. Quantum emitters have been extensively studied over the past decade, emerging as a leading platform for quantum sensing \cite{wang2025cavity,barry2020sensitivity,wang2024spin} and quantum networking~\cite{humphreys2018deterministic, Duan:21}. Previous experiments have achieved resolution around
$\sigma \sim 0.7$~nm due to scaling limits inherent in conventional localization methods \cite{chen2019superresolution,rendler2017optical,bersin2019individual}. In this work, we apply widefield photoluminescence excitation (PLE) with DIGIT to overcome the shot-noise scaling of localization with photon counts, achieving an unprecedented localization precision of 0.178~\AA. We further demonstrate that this method can be extended to large-scale emitter arrays, enabling massively parallel localization for different emitter clusters. These results show that DIGIT unlocks a potential avenue to applications ranging from identifying solid-state quantum memories in crystals to the direct observation of optical transitions in the electronic structure of molecules. 

\section{Results}

\subsection{DIGIT concept}

We introduce the DIGIT in the context of single-molecule localization microscopy (SMLM). SMLM is one of the most powerful super-resolution modalities, enabling the visualization of emitters down to the molecular level~\cite{reinhardt2023aangstrom}. In conventional SMLM, each emitter produces a diffraction-limited point-spread function PSF, whose centroid $(\mu_x, \mu_y)$ is estimated with precision $\sigma$, fundamentally limited by $\sigma \geq \sigma_0 / \sqrt{N}$, where $\sigma_0$ is the standard deviation of the PSF and $N$ is the total photon count~\cite{SMIM_2021} (Fig.~\ref{fig1}a). After multiple SMLM measurements, we obtain a continuous probability density function of emitter locations $f(\mu_x)$, as shown in Fig.~\ref{fig1}b.  In the case of solid-state crystal, however, emitters are only allowed to sit at discrete lattice sites identified by the substitute carbon atoms (Fig.~\ref{fig1}c). In our implementation, we account for this by using a Bayesian statistical process, where we update the initial probability distribution $f(\mu_x)$ to the posterior localization $P(x)$ given this underlying atomic lattice structure $\delta(x-na)$, where $a$ is the lattice constant. This posterior collapses onto sharp peaks at $x = na$, as depicted in Fig.~\ref{fig1}d. This concept naturally extends to two dimensions as shown in Fig.~\ref{fig1}e, where each emitter position is mapped onto the lattice grid as a discrete integer combination of lattice vectors: $\boldsymbol{\mu}_i = m_i \mathbf{a}_1 + n_i \mathbf{a}_2$, with $(\mathbf{a}_1, \mathbf{a}_2)$ denoting lattice unit vectors and $(m_j, n_j) \in \mathbb{Z}$. The color-coded SMLM localization illustrates how DIGIT refines each localization onto the lattice. The resulting DIGIT localization $\sigma_p$ is shown in Fig.~\ref{fig1}f as a function of $N$. Compared with the conventional position estimation, $\sigma_p$ represented in the green line scales $\sigma_p\propto e^{-\sqrt{N}}$, diminishing exponentially after $\sigma$ surpasses the lattice constant $a$, achieving atomic localization.

\begin{figure*}
\includegraphics[width=0.9\textwidth]{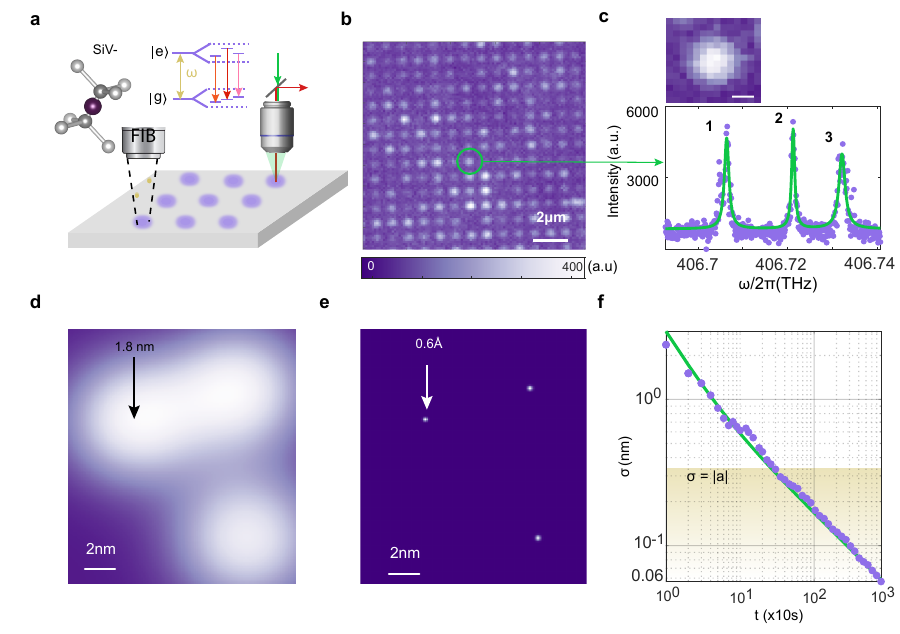}
\caption{\label{fig2} \textbf{Sub-diffraction imaging of SiVs.} \textbf{a,} Schematic of sample preparation and characterization. SiVs are created via FIB implantation. The quantum emitters are optically excited using a 532 nm green laser, and the resulting PL is collected through a confocal microscope operated at cryogenic temperatures. Inset: Energy level structure of SiV centers. The optical transition frequency $\omega$ exhibits inhomogeneous broadening across different emitters, enabling SMLM with PLE spectroscopy.
\textbf{b,} Diffraction-limited PL image of implanted SiV clusters. \textbf{c,} Top: EMCCD image of one resolved emitter. Scale bar: $200$~nm. Down: PLE spectrum of the cluster. Three emitters separate in frequency space within the diffraction-limited spot, with Lorentzian fitting shown in green. \textbf{d,}
\textbf{e,} Comparison of SMLM-reconstructed images without and with digital twin calibration. In \textbf{d,} closely spaced emitters appear unresolved, whereas in \textbf{e,} distinct emitters emerge with a localization uncertainty towards $\sigma = 0.6$~\AA. \textbf{f,} The localization precision $\sigma$ improves following the photon-shot-noise-limited scaling with an SMLM technique.}
\end{figure*}

\subsection{SMLM for a quantum emitter}
A prerequisite for implementing DIGIT is achieving localization accuracy $\sigma$ comparable to the lattice constant $a$. Here, we adapted PLE to achieve lattice-scale accuracy and precision. Compared with other emitter localization methods like stochastic optical reconstruction
microscopy \cite{pfender_single-spin_2014,chen_nanoletter_2013}, structured illumination microscopy \cite{ShanAPL2024, rittweger2009sted,ArroyoNanoLetter2013} and quantum state control \cite{ChenPRAppl2019, HuangPRA2020}, PLE allows deterministic excitation control of individual targeted emitters in a dense cluster by spectrum separation due to the inhomogeneous distribution nature of quantum emitters. 


We use silicon-vacancy (SiV) centers as exemplary quantum emitters due to their exceptional stability and resistance to photobleaching, making them well-suited for repeated or prolonged imaging applications. Moreover, the mature focused ion beam (FIB) enables nanometer-scale resolution in the implantation process. Despite the high spatial precision, the overall number of implanted centers can be kept low, which helps to maintain well-isolated emission sites and minimizes background interference.


We first use PLE spectroscopy to spectrally resolve individual emitters. As shown in Fig. \ref{fig2}a, a SiV array is implanted in diamond with FIB and imaged with a widefield cryogenic microscope. A representative photoluminescence (PL) image is displayed in Fig. \ref{fig2}b, where each bright spot may contain multiple SiVs separated by sub-diffraction-limit distances. To resolve these closely spaced emitters individually, we sweep the excitation laser frequency across the zero-phonon-line transitions while simultaneously recording emission into the phonon sideband using an electron-multiplying charge-coupled device (EMCCD) camera. As shown in Fig. \ref{fig2}c, three distinct spectral peaks at frequencies $\omega_1,~\omega_2,~\omega_3$ correspond to three individual emitters within the cluster shown in Fig. \ref{fig2}b.

\begin{figure*}
\includegraphics[width=1\textwidth]{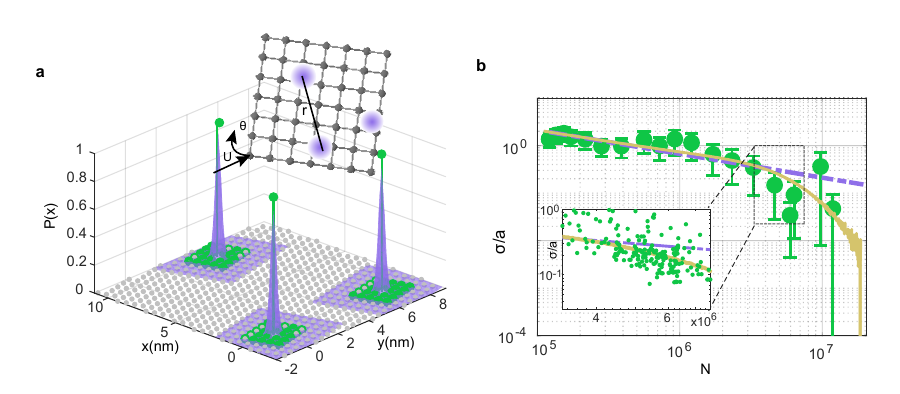}
\caption{\label{fig3} \textbf{Atomic-scale precision achieved by DIGIT.} \textbf{a}, Comparison of SMLM localization distribution $f(x)$ in purple versus DIGIT posterior distribution $P(x)$ in green, aligned along the lattice by grey dots. The inset illustrates the diamond lattice model projected along the [100] crystallographic direction, including lattice's global rotation $\theta$ and offset $U$. \textbf{b}, Normalized localization precision $\sigma$ distance between all emitter pairs as a function of total collected number of photons $N$. The experimentally measured localization precision using DIGIT (green) with its simulation predictions (yellow), demonstrating an exponential improvement over the SMLM shot noise scaling (purple). The inset zooms in the transition where DIGIT localization begins to surpass conventional SMLM.}
\end{figure*}

We then sequentially localize these three emitters with multiple localizations. The initial localization uncertainty fitted by a 2D Gaussian distribution yields $\sigma = 1.85$~nm over $8000$ frames, as shown in Fig.~\ref{fig2}d. This precision is insufficient to resolve the individual emitters due to the inherent inaccuracies in the Gaussian approximation for the collected PSFs. To address this limitation, we further apply established calibration and correction techniques tailored to the emitters' system. These include EMCCD inter-pixel gain correction, optical aberration correction through adaptive optics and PSF modeling, dipole-orientation characterization, and drift correction via Kalman filter~\cite{pertsinidis2010subnanometre, backlund2016removing}. These improvements enable us to achieve a localization precision $\sigma=0.6$~\AA \ as illustrated in Fig.~\ref{fig2}e. The localization precision follows the shot-noise-limited scaling (Fig. \ref{fig2}f) \cite{mortensen2010optimized} and surpasses the threshold at the diamond lattice constant $a$ when $t>300~$s.

\subsection{Atomic Localization with DIGIT}

\begin{figure*}
\includegraphics[width=1\textwidth]{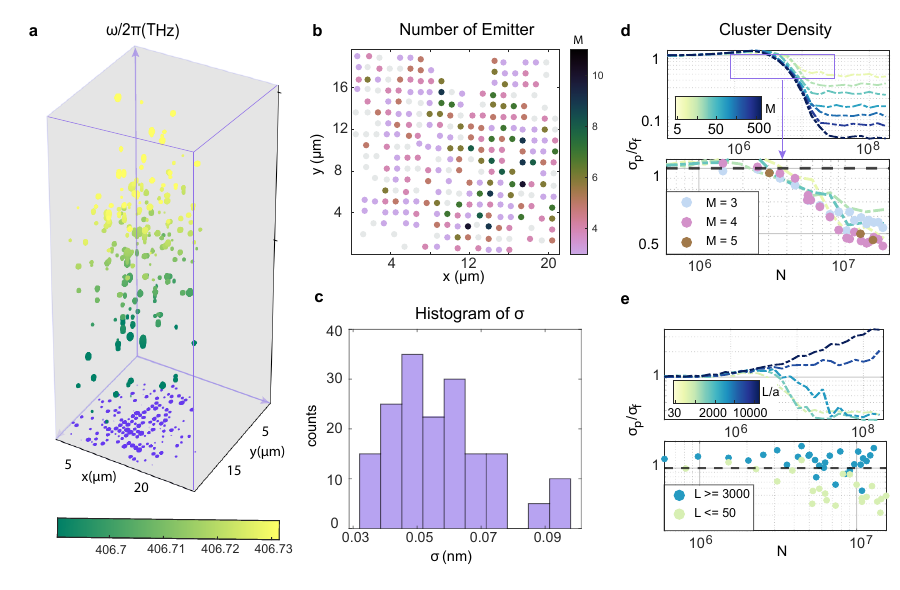}
\caption{\label{fig4}\textbf{Large-scale DIGIT.} \textbf{a,} Three-dimensional representation of widefield PLE obtained by sweeping a resonant laser over 30~GHz in a $10$~MHz increment. \textbf{b,} Number of emitters in each reconstructed cluster. \textbf{c,} Histogram of the measured precision $\sigma$ for individually resolved emitters. \textbf{d,} Top: Numerical simulations of the precision ratio $\sigma_p/\sigma_f$ as a function of the number of measured photons $N$ for different numbers of emitters $M$. Bottom: Representative experimental data with theoretical fits for $M = 3, 4, 5$. Dashed line: $\sigma_p/\sigma_f = 1$. \textbf{e,} Top: Numerical simulations of the precision ratio $\sigma_p/\sigma_f$ as a function of the number of measured photons $N$ for various cluster sizes $L$, with $M = 5$ fixed. Bottom: Representative experimental data with theoretical fits showing the dependence on $L\geq 3000$ and $L\leq50$. Dashed line: $\sigma_p/\sigma_f = 1$.}
\end{figure*}

Building upon the localization results, we applied DIGIT to map the positions of SiVs, encoding their locations with respect to the diamond lattice constraints. Although the full three-dimensional diamond lattice is inherently complex, we simplify it by projecting the lattice along the [100] direction of the diamond substrate $\delta(x,y)$~\cite{hu2024developing}, as shown in Fig.~\ref{fig3}a. To further enhance the accuracy of the Bayesian model used in DIGIT, we mapped the measured coordinates to lattice sites through an affine transformation. This transformation accounts for the global rotation $\theta$ and offset of the lattice $U$, aligning the experimental data with the periodic structure of the lattice. The parameters for this transformation were optimized using maximum-likelihood estimation, as detailed in Extended Data Fig.~3. The resulting DIGIT localization of the three emitter clusters is shown in Fig.~\ref{fig3}a, where the conventional SMLM localization distribution breaks into discrete peaks that align with the fitted lattice sites, significantly enhancing the localization probability at certain lattice sites.

We then analyze the localization scaling as a function of $N$. With a conventional SMLM, the localization is proportional to the square root of the measured photon number $N$, as shown of the purple line in Fig.~\ref{fig3}b. By applying the concept of DIGIT, when the precision $\sigma$ is much larger than the lattice constant ($\sigma\gg a$), the precision follows the same trajectory as the conventional SMLM. However, when $\sigma/a$ approaches $0.33$, the localization scaling deviates from the conventional scaling and drops exponentially with the number of measured photons, as depicted in the yellow curve. We then demonstrate this concept with experimental data. As shown in Fig. 3b, the experimental results of the emitter localization fit well with the modeling and demonstrate an unprecedented localization of $\sigma_p = 0.178 \pm 0.107$~\AA~below the diamond lattice spacing, while conventional SMLM yields a mean uncertainty of $\sigma_f = 0.7$~\AA~for the same three emitter pairs.


\subsection{Large-scale DIGIT}


Although scanning-based approaches such as stimulated emission depletion microscopy and MINFLUX have achieved lateral resolutions in the single-digit nanometer range \cite{scheiderer2025minflux,gwosch2020minflux,vicidomini2018sted}, their point-by-point acquisition inherently limits scalability. In contrast, DIGIT enables massively parallel detection, and here we experimentally demonstrate this capability to simultaneously resolve multiple emitter clusters across a $20 \times 20$~$\mu$m field of view.


We present a hyperspectral image of widefield PLE, spanning spatial $(x, y)$ and atomic transition frequency $\omega/2\pi$, as shown in Fig.~\ref{fig4}a. We applied intensity-based region grouping to the PLE EMCCD image to distinguish individual emitters whose transitions range from $406.69$ to $406.73$~THz. We successfully identify the number of emitters $M$ at each sub-diffraction limit spot, as shown in Fig.~\ref{fig4}b. A detailed flowchart of the widefield PLE data analysis can be found in Extended Data Fig.~4. 


We then integrated $8000$ frames at each resolved frequency $\omega_i$ and the measured localization uncertainty $\sigma$ for each emitter is summarized in Fig.~\ref{fig4}c. Only SiVs exhibiting sub-\AA ngstr{\"o}m $\sigma$ were counted for subsequent analysis. Emitters that did not meet this requirement have typically reduced brightness, exhibit blinking behavior \cite{Hepp2014}, or experience non-uniformity widefield illumination \cite{sutula2023large}, all of which contribute to worse localization uncertainty. As shown in Fig.~\ref{fig4}c, we achieved sub-\AA ngstr{\"o}m localization in $172$ emitters, with a mean localization precision of $0.05$~nm. Those results represent the first demonstration of massively parallel imaging with sub-\AA ngstr{\"o}m precision, enabled by the unique combination of wide-field imaging and DIGIT’s ability to resolve clustered emitters onto discrete lattice sites, thereby enhancing localization probability and precision with large-scale emitters. 

Large-scale DIGIT enables the exploration of localization dependence across different clusters in parallel. We first analyze the localization enhancement, defined as $\sigma_p/\sigma_f$, at various clusters with differing emitter densities. The presence of additional emitters provides improved lattice alignment information through maximum likelihood estimation, resulting in a localization enhancement that scales as $\sigma_p/\sigma_f \propto 1/M$, as shown in Fig.~\ref{fig4}d. We validate this prediction using experimental data. Specifically, we examined three cluster densities, $M = 3, 4,$ and $5$. Compared to the previous case in Fig.~\ref{fig3}, increasing $M$ beyond 3 yields a clear improvement in localization. However, no significant difference was observed between the cases with $M = 4$ and $M = 5$, which may be attributed to the limited number of clusters with $M = 5$ under our FIB conditions and potential lattice damage during the emitter generation process.

The spatial separation $L$ is another key factor in the localization process. When emitters within a cluster are further apart, the localization enhancement provided by DIGIT diminishes. This reduction in performance arises because lattice fitting errors become dominant when distinguishing the maximum likelihood lattice point from adjacent points. In such cases, increasingly precise estimates of the rotation angle $\theta$ are required to satisfy the condition $L\theta \leq a$.
We examine the ratio $\sigma_p/\sigma_f$ for clusters with varying spatial separations at $M = 5$. As shown in the top panel of Fig. \ref{fig4}e, there exists a critical separation $L = L_0$ at which DIGIT can provide enhancement. For the current parameter set, we theoretically identify $L_0 = 3000a$. Experimental results for $L \geq 3000a$ and $L \leq 50a$ are presented in the bottom panel of Fig. \ref{fig4}e. The average value of $\sigma_p/\sigma_f$ clearly separates around 1 under these two specific conditions, aligning well with theoretical predictions. Moreover, in experimental settings, this effect is further complicated by strain-induced variations in the lattice coherence length, which can result in uneven lattice structures and discontinuities that hinder precise localization \cite{tong2024structural}.








\section{Discussion and Outlook}
As the localization precision of conventional super-resolution microscopy approaches the length scale of electronic orbitals coupled to optical fields, it becomes essential to incorporate the atomic structure of samples as a Bayesian prior. Our demonstration of the DIGIT was made possible by achieving localization precision ($\sigma$) comparable to the lattice constant ($a$), allowing emitters to be accurately mapped to discrete lattice points rather than being confined to a continuous spatial domain. Furthermore, DIGIT provides a new precision scaling law: $\sigma_p \propto e^{-\sqrt N}$ that is isotropic in all dimensions. Besides, parallelized DIGIT demonstrates its scalability by simultaneously localizing multiple emitters. To further access denser clusters, improved spectral resolution becomes critical. One promising strategy involves operating PLE without a repump laser so that emitters approach lifetime-limited spontaneous emission~\cite{Arjona2022, ikeda2024}. 

Looking ahead, DIGIT has a direct application in diamond-based quantum information processes. For instance, the ability to achieve sub-\AA ngstr{\"o}m localization enables in-situ tracking of emitters' diffusion and interaction~\cite{zu2021emergent,Mosavian2024,rendler2017optical,Xie2022}, increases spatial storage density~\cite{zhou2024terabit}, and facilitates the generation and control of large cluster states made of locally coupled spins~\cite{Nemoto2014,maurer2010far}. Additionally, it is readily transferable to other periodic host materials, aiding to identify molecular-scale configurations. Potential candidates include point-like defects~\cite{Viktor2017, koehl2011room, stern2022room}; extended electronic orbitals in semiconductor quantum dots \cite{englund2007controlling, srivastava2015optically}, or molecules such as dibenzoterrylene, tetracene \cite{Wu2014, einzinger2019sensitization, akselrod2014visualization, Musavinezhad2024, Laorenza2021} and nuclear pore complex~\cite{Hampoelz2019,Hoelz2011}. 

However, DIGIT is currently constrained to applications with well-characterized lattice structures. To overcome this limitation, our theoretical proposal for diamond suggests that with higher density CC, the atomic structure and the associated electronic orbital transitions can be learned, all from optical observation with zero-knowledge base (see Supplementary Information Sec. VI). Advanced computational techniques, such as deep learning of PSF from experimental observations~\cite{Mockl:20, cao2024neural} and surrogate models~\cite{rodriguez2024automated}, or generative AI methods such as adaptive model-based measurements, promise further acceleration.

\section*{Acknowledgment}

We thank M.Mazaheri, Q.Peng for valuable input with the digital twin localization model. We thank C.S. Peng, A. Roy, J. Ren, Y. Liu for helpful discussions. We thank M.Mazaheri, V.Saggio and X. Chen for proofreading the manuscript. We acknowledge excellent technical support at the Massachusetts Institute of Technology (MIT) for MIT.Nano. Y.Yu performed FIB implantation, J. Daley performed evaporation metal growth. This work was supported in part by the National Science Foundation (NSF) Science-Technology Center (STC) Center for Integrated Quantum Materials (CIQM) under Grant No. DMR-1231319, by the NSF Engineering Research Center for Quantum Networks (CQN) awarded under Cooperative Agreement No. 1941583, and by the MITRE Moonshot Program. Y.D. acknowledges support by MathWorks fellowship.

\section*{Author Contributions}
Y.D. conducted diamond fabrication, SiV implantation, digital twin development, localization analysis and measurements of the whole experiment. K.C.C. assisted with diamond fabrication. M.E.T assisted with building the wide-field SMLM, the dipole radiation model and digital twin verification. Q.G, H.W, Y.H, and D.R.E assisted with calibrating the camera photoresponse. Y.D and Q.G developed and implemented the Bayesian framework. Y.D, M.E.T and D.R.E conducted the theory proposal of optically resolved lattice structure. Y.D and D.R.E. conceived and supported the project. Y.D., M.E.T. and D.R.E wrote the manuscript with input from all authors.


\section*{Data availability}
All the data that support the findings of this study are included in the Article and its Supplementary Information. Source data are available via figshare at \hyperref[10.6084/m9.figshare.28281212]{10.6084/m9.figshare.28281212}

\section*{Code availability} 
Code for DIGIT principle and widefield DIGIT is available at: \hyperref[https://github.com/sophiaOnPoint/DIGIT]{https://github.com/sophiaOnPoint/DIGIT}



\bibliography{apssamp}

\section{Method}
\subsection{Diamond fabrication}
We first strain-relieved the surface of an electronic-grade, single-crystal diamond plate (Element 6) through plasma etching \cite{Burek2012}. Next, we deposited a $180$~nm layer of silicon nitride (SiN) hard mask via plasma-enhanced chemical vapor deposition (PECVD). The SiN hard mask was then patterned with coordinate markers by electron-beam lithography and $\text{CF}_4$ reactive ion etching. Subsequently, we employed inductively coupled reactive-ion etching (ICP-RIE) to transfer the pattern from the hard mask onto the diamond surface. After etching the diamond with oxygen plasma, we deposited a $10$~nm gold layer using thermal evaporation, followed by a lift-off procedure that removes the SiN mask.
\subsection{Ion implantation}
Focused ion implantation was performed at the MIT Nano facility using FIB-SEM VELION (Raith). It is a FIB machine using an ExB filter and single ion implantation, using fast beam blanking. The ExB mass-filter separates out different ionic species and charge states from liquid metal alloy ion sources, providing the capability for implantation of both Si and Ge at 35 keV and 70 keV. For the Si implantation discussed here, we used a GeSiAu liquid metal alloy ion source with typical Si beam currents ranging from 0.4 to 1~pA. Fast beam blanking allows direct control over the number of implanted ions. We determine the number of implanted ions by measuring the beam current and setting the pulse length to target a given number of ions per pulse. We used an effective areal dose of 75, 70 and 105 ions/spot for Si++ implantation at 70~keV with $6$~nm beam diameter. Stopping and Range of Ions in Matter (SRIM) simulations estimated an implantation depth of $50\pm15$ nm. After ion implantation, we annealed the sample at $1050$~$^{\circ}\mathrm{C}$ under high vacuum ($<10^{-6}$~mbar) for 2~h to form SiVs and eliminate other multi-vacancy defects. Finally, we submerged the diamond sample in boiling tri-acid treatment (1:1:1 nitric:perchloric:sulfuric) to remove residual graphite induced from annealing, and subsequently oxidized the surface in a $30\%$ oxygen atmosphere at $450$~$^{\circ}\mathrm{C}$ for 4~h \cite{Fu2010}.

\subsection{Experimental Setup}
The PLE SMLM measurements were performed at 4~K in a closed-cycle helium cryostat (Montana Instruments). A home-built wide-field microscope (Supplementary Fig.~1) collected the CCs' fluorescence with a high-NA objective (Zeiss 100$\times$. NA 0.95) and directed the emission to either an EMCCD (Cascade 1K) with a $f=100$~mm imaging lens or a single-mode fibre coupled to both an avalanche photodiode (PerkinElmer) and a spectrometer (Princeton Instruments). The effective pixel size is $75$~nm (after 1$\times$1 binning). Super-resolution PLE was performed using a tunable laser (MSquared SolsTis) and a 532-nm green laser for charge state initialization. The resonant laser output passed through a bandpass filter (Semrock FF01-735/28) to minimize background fluorescence. The CCs' emission was collected via a dichroic mirror and filtered in free space using bandpass filters (Semrock FF01-740/13). A three-axis closed-loop piezoelectric stage (Attocube ANC300) was used to control the position of the sample on the microscope stage. 

\subsection{Widefield PLE}
We performed widefield PLE of SiV centers by using acousto-optic modulators to deliver a resonant pulse and a 532~nm charge repump pulse. The resonant laser was swept across a $30$~GHz range in $10$~MHz increments, with its frequency stabilized via a HighFinesse WS7 wavemeter to ensure high spectral resolution. We optimized the signal-to-noise ratio (SNR) of SiV EMCCD imaging by choosing the following camera's setting: acquisition time at $t = 1.2$~s, binning = 1$\times$1, electron amplification of $1200$ in addition to the camera's gain of $3$.

\subsection{DIGIT Algorithm}
We evaluate the performance of DIGIT with the following step-by-step algorithm:

(1) A grid of $L$ lattice sites with $M$ number of emitters randomly positioned is generated (ground truth).

(2) SMLM localization $f(\mu)$ is modeled from a 2D Gaussian distribution with linewidth $\sigma$.

(3) Prior knowledge of lattice structure is modeled as a series of delta functions $\delta(x,y)$ with spacing aligned with lattice periodicity.
\begin{equation*}
    Pr(x) = \sum_{n\in\mathbb{Z}}\delta(x-n)
\end{equation*}

(4) Using MLE to find the optimized lattice offset $U$ and rotation angle $\theta$ based on the SMLM localization.

(5) Posterior $P(x,y)$ is calculated in the Bayesian framework based on likelihood $f(\mu)$ and prior $\delta(x,y)$. The result is:
\begin{equation*}
    P(x,y|\mu_x,\mu_y) = \frac{1}{K}\frac{e^{-(\frac{x-\mu_x}{\sqrt{2}\sigma})^2}e^{-(\frac{y-\mu_y}{\sqrt{2}\sigma})^2}}{\sum_{m,n\in\textit{Z}}e^{-(\frac{m-\mu_x}{\sqrt{2}\sigma})^2}e^{-(\frac{n-\mu_y}{\sqrt{2}\sigma})^2}}
\end{equation*}
where $\sum P(x,y) = 1$. 

(6) DIGIT localization precision is calculated by finding the standard deviation of $P(x,y)$.

Steps 1 to 5 are repeated for different values of $\sigma$, $M$ and $L$ to numerically study $\delta(x,y)$.

\subsection{DIGIT Localization}
Extended Data Fig.~2 shows the digital-physical twin system we developed to localize the center of the EMCCD image. A digital twin is a virtual representation of a physical object that is updated with real-time process. Compared with a model fitting, the digital twin enriched it by using a live feedback to optimize the setup's performance~\cite{Kenett2022}. 

\subsubsection{Digital Twin}
We constructed the digital twin starting from a dipole radiation model adapted from ref.~\cite{ Axelrod2012}. We included the following parameters to analytically calculate the dipole's farfield emission intensity $E(x,y)$: emitter's position $(x,y,z)$~nm, emission polarization $(\theta,\phi)$, system defocus and astigmatism, objective numerical aperture (NA), focal length, and working distance. 

We further added a matrix $[\gamma]$ that corrects environmental imperfection to $E(x,y)$. We used a calibration sample with 1x1~$\mu$m grid pattern (EM-Tech M-1) to extract the conversion factor from pixel to absolute distance. We used the python package [distortion] to fit the grid pattern to correct for distortion in the EMCCD images. The full calibration details are covered in the Supplementary Information Sec. III.

Another significant correction concerned setup drift during measurements. We addressed this by implementing a Kalman filter to track and correct the drift trajectory over time. The Kalman filter is based on a state model (Gauss-Markov model), which characterizes the dynamics of gain and bias, and an observation model that minimizes the estimation error of the state position. The detailed algorithm is provided in the Supplementary Information Sec. III. We evaluated the effectiveness of our drift correction approach using Allan variance.

The digital twin further optimized imaging performance via a custom-built control system. Specifically, the system automatically adjusts individual mirrors to align both the tunable and green lasers so that their focal spots coincide according to the cost function $\mathrm{Conv}(I_{\mathrm{green}}, I_{\mathrm{resonant}})$. In addition, we minimized defocus by monitoring the PSF intensity and adjusting the piezo stage accordingly.

\subsubsection{Localization}
We first corrected the EMCCD images based on the interpixel calibration, aberration and drift analysis, then we fitted the dipole radiation model based on a weighted MLE cost function to localize emitters' center $(x_i, y_i)$:
\begin{equation*}
    (x_i,y_i) \rightarrow \text{argmin}\left[\frac{1}{w_i}(E(x_i,y_i,\phi) - I(x,y))\right]
\end{equation*}
where $E(x_i,y_i,\phi)$ is the dipole radiation model, $I(x,y)$ is the corrected EMCCD image, and the weight $w_i$ is square root of pixel intensity $1/\sqrt{I}$ to balance the information at the emission center and tail. We repeat this fitting for all independent measurements, and the theoretical SMLM bound of localization precision is:
\begin{equation*}
    \langle\delta_x\rangle^2 = \frac{\sigma_0^2+a^2/12}{N} + \frac{8\pi \sigma_0^4b^2}{a^2N^2}
\end{equation*}
where $a$ is the pixelization, $b$ is readout noise, $N$ is photon counts of the EMCCD image.

\subsection{Large-scale DIGIT Analysis}
Extended Data Fig.~4 illustrates the workflow for localizing emitters across multiple clusters using DIGIT. First, we performed widefield PLE imaging and summed all frequency-step images to generate a single composite. A 2D Gaussian bandpass filter was then applied to enhance the SNR, with low-pass and high-pass cutoffs set by $\sigma_L$ and $\sigma_H$, respectively. This filtered image was used to identify candidate emitter centers $(x_i,y_i)$ where each bright spot must have an apparent size within $\sigma_L$ and $\sigma_H$ (Supplementary Information Sec. V). 

Having identified these candidate emitters, we defined a region of interest around each bright spot. We then extracted the fluorescence spectrum over the full frequency range and fitted each peak with a Lorentzian lineshape. Supplementary Fig.~11 shows examples of the PLE spectrum. Subsequently, our custom control software locked the laser to these resonance frequencies, and we re-acquired the corresponding PSFs for each emitter.

We then performed a least-squares fit to a 2D Gaussian model of each PSF. If the initial fit achieved an R-squared $r^2 > 0.6$, we proceeded with the DIGIT-based (digital twin) fitting to refine the localization accuracy.

\end{document}